\documentclass{article}

\usepackage{microtype}
\usepackage{graphicx}
\usepackage{subfigure}
\usepackage{booktabs} 
\usepackage{graphicx} 
\usepackage{amsmath}
\usepackage{algorithm}
\usepackage{algorithmic}
\usepackage{xcolor}
\usepackage[colorlinks=true, linkcolor=blue, citecolor=blue, urlcolor=blue]{hyperref}
\usepackage{amsfonts}
\usepackage{amssymb}
\usepackage{amsmath}
\usepackage{graphicx}

\usepackage{float}
\usepackage{tikz}
\usetikzlibrary{arrows.meta, positioning, calc}
\newcommand{\RETURN}{\STATE \textbf{return}}

\newenvironment{breakablealgorithm}
  {\begin{center}
   \refstepcounter{algorithm}
   \hrule height.8pt depth0pt \kern2pt
   \renewcommand{\caption}[2][\relax]{%
     {\raggedright\textbf{Algorithm \thealgorithm} ##2\par}%
     \ifx\relax##1\relax 
       \addcontentsline{loa}{algorithm}{\protect\numberline{\thealgorithm}##2}%
     \else 
       \addcontentsline{loa}{algorithm}{\protect\numberline{\thealgorithm}##1}%
     \fi
     \kern2pt\hrule\kern2pt
   }
  }
  {\kern2pt\hrule\relax\end{center}}

\usepackage{hyperref}

\usepackage[accepted]{mlsys2025}

\usepackage{eso-pic}
\usepackage{xparse}

\newlength{\badgewidth}
\newlength{\badgegap}
\newcommand{\badgeList}{}

\NewDocumentCommand{\addTopRightBadge}{O{} m}{%
\gappto{\badgeList}{\href{#1}{\includegraphics[width=\badgewidth]{#2}}\hspace{\badgegap}}%
}

\newcommand{\placeTopRightBadges}{%
\AddToShipoutPictureBG*{%
\put(\LenToUnit{\paperwidth - 1.5cm - \badgewidth},\LenToUnit{\paperheight - 2cm}){%
\makebox[0pt][r]{\badgeList}%
}%
}%
}

\addTopRightBadge{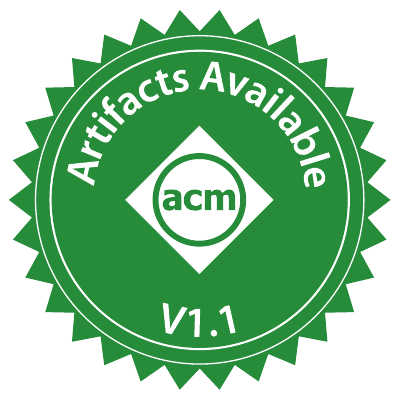}
\addTopRightBadge{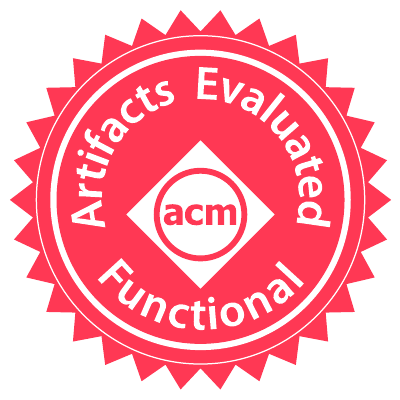}
\addTopRightBadge{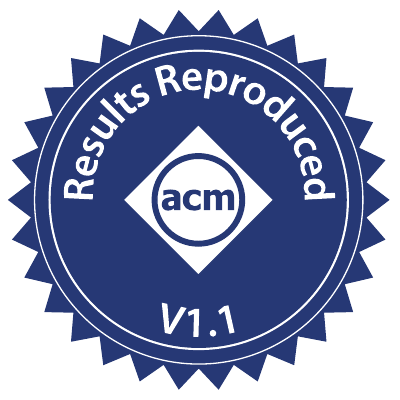}

\placeTopRightBadges

\mlsystitlerunning{Hawkeye: Reproducing GPU-Level Non-Determinism}

\begin{document}

\twocolumn[
\mlsystitle{Hawkeye: Reproducing GPU-Level Non-Determinism}



\mlsyssetsymbol{equal}{*}

\begin{mlsysauthorlist}
\mlsysauthor{Erez Badash}{equal,prl}
\mlsysauthor{Dan Boneh}{equal,sto}
\mlsysauthor{Ilan Komargodski}{equal,prl}
\mlsysauthor{Megha Srivastava}{equal,sto}
\end{mlsysauthorlist}

\mlsysaffiliation{prl}{Pearl Research Labs}
\mlsysaffiliation{sto}{Department of Computer Science, Stanford University}
\mlsyscorrespondingauthor{Erez Badash}{erez.badash@pearlresearch.ai}
\mlsyscorrespondingauthor{Megha Srivastava}{megha@cs.stanford.edu}
\mlsyscorrespondingauthor{Dan Boneh}{dabo@cs.stanford.edu}
\mlsyscorrespondingauthor{Ilan Komargodski}{ilan.omargodski@pearlresearch.ai}

\mlsyskeywords{Verifiable Machine Learning, Nondeterminism}

\vskip 0.3in

\begin{abstract}
We present Hawkeye, a system for analyzing and reproducing GPU-level arithmetic operations. Using our framework, anyone can re-execute on a CPU the exact matrix multiplication operations underlying a machine learning model training or inference workflow that was executed on an NVIDIA GPU, without any precision loss. This is in stark contrast to prior approaches to verifiable machine learning, which either introduce significant computation overhead to the original model owner, or suffer from non-robustness and quality degradation. 
The main technical contribution of Hawkeye is a systematic sequence of carefully crafted tests that study rounding direction, subnormal number handling, and order of (non-associative) accumulation during matrix multiplication on NVIDIA's Tensor Cores.  We test and evaluate our framework on multiple NVIDIA GPU architectures ( Ampere, Hopper, and Lovelace) and precision types (FP16, BFP16, FP8). In all test cases,  Hawkeye enables perfect reproduction of matrix multiplication on a CPU, paving the way for efficient and trustworthy third-party auditing of ML model training and inference.  We provide source code for Hawkeye  \href{https://github.com/badasherez/gpu-simulator}{here}.
\end{abstract}
]



\printAffiliationsAndNotice{\mlsysEqualContribution} 

\section{Introduction}

Modern machine learning (ML), including both model training and inference, has become increasingly compute-intensive, driven by the rapid growth in number of model parameters, dataset size, and architectural complexity \cite{Kaplan2020ScalingLaws,Hoffmann2022Chinchilla}. This is especially true with large language models, and the
training and inference of state-of-the-art models now often requires massive compute clusters and careful orchestration of distributed systems \cite{Shoeybi2019MegatronLM,Rajbhandari2020ZeRO}.
To meet these growing demands, a new class of training and inference platforms has emerged, offering ML-as-a-service to offload the heavy computational burden from end users.
For instance, cloud-based solutions like AWS SageMaker \cite{AWSSageMakerDocs}, Google Vertex AI \cite{VertexAIDocs}, and Azure ML \cite{AzureMLDocs} provide full-stack infrastructure for model training, tuning, and deployment, while companies such as Replicate \cite{ReplicateDocs} and TogetherAI \cite{TogetherAIDocs} provide API endpoints to support clients who lack the resources to train a model themselves. 

However, these services require clients to trust them to train or run inference correctly, without introducing any backdoors, shortcuts (e.g. training for a reduced number of steps), or other runtime modifications to the provided models. \textit{How can we guarantee that the service provider executed the task as specified, and hold them accountable?} This task, \textbf{verifiable ML}, seeks to design ML training and inference platforms with the additional guarantee that all tasks are executed correctly ~\cite{Peng2025SurveyZKML, Sun2024zkLLM, Ganescu2024TrustTheProcess, Lycklama2024Artemis, SrivastavaAB24, So2024oppAI}.

There has been a long history in the cryptography and security literature on verifiable computation,  where the typical setting considers a powerful server that performs a heavy computation for a computationally weak verifier \cite{Shamir1992IPPSPACE, Kilian1992EfficientZK, GKR2008Delegating, GGPR2013QSP, PHGR2013Pinocchio, Groth2016, Maller2019Sonic, Gabizon2019Plonk, Chiesa2020Marlin, BCH+2018STARKs, Aurora2019, Chiesa2019Fractal, Setty2020Spartan, Hyrax2018, Libra2019, KZG2010, FRI2018, Bulletproofs2018}. These techniques guarantee that the prover executes the prescribed computation, without requiring the client to fully re-execute it. However, a prerequisite of these works is that the computation performed by the server is deterministic, which is not true for modern ML workflows \cite{hutson2018artificial}. 

Typical ML training and inference tasks  heavily depend on software and hardware implementations, which oftentimes are not fully known (see Section~\ref{sec:challenge} for details). A primary source of non-determinism is in the accumulation process within specialized hardware units, such as the widely used NVIDIA Tensor Cores.\footnote{Tensor Cores are specialized hardware units that perform highly optimized matrix multiplications.} Non-determinism arises from many unspecified details including rounding strategy, subnormal numbers, and the order of accumulation during matrix multiplication operations (as floating point arithmetic is not associative)~\cite{DBLP:conf/sc/ShanmugaveluTCH24}.  Thus, the same model provided the same input and random seed on different GPUs could result with different inference outputs (see \cite{SrivastavaAB24} for an example). 

\paragraph{Motivation.} 
Suppose an auditor wishes to \emph{re-execute} a specific computation on a CPU in order to verify correctness. For example, an auditor may wish to verify that a service provider hosting a language model is indeed providing the correct next-token output during inference, with respect to a particular sampling method.  In an idealized world of purely deterministic computation, such a re-execution would either match the outputs exactly, or reveal discrepancies attributable to incorrect execution (e.g., a service provider wishing to cut costs by serving a smaller language model). However, in practice, running the same computation on different hardware, or even on the same hardware with different low-level execution schedules, can produce  different numerical results.  

Consider the following concrete simplified FP16 vector-vector multiplication $A\cdot B^T$ (which can easily be generalized to matrix-matrix multiplication):
\[
A =
\begin{bmatrix}
65504 & 1 & -65504 & 1
\end{bmatrix},
\]\[
B=\begin{bmatrix}
1 & 0.001 & 1 & 0.001
\end{bmatrix}.
\]
On Tensor Cores of NVIDIA's L40S GPU (Ada Lovelace architecture), the result is~$0$, whereas on NVIDIA's A100 GPU (Ampere architecture), the result is~$0.0020$. Such small discrepancies can accumulate over the course of an entire repeated inference or training run, resulting in different model weights and downstream predictive behavior. Additional examples on the effect of non-associativity on the stability of computations, or downstream performance of models, can be found in~\cite{DBLP:conf/sc/ShanmugaveluTCH24, SrivastavaAB24}.

As a result, simply comparing outputs between the original execution and an auditor's re-execution is insufficient for verification: even a perfectly correct computation may not result in a bitwise match. 
To \emph{provably} avoid ambiguity in ML computations, one systematic approach is to disable non-deterministic hardware features. This guarantees reproducibility, but often incurs significant slowdowns~\cite{ArunArnaud25}. An alternative  approach~\cite{SrivastavaAB24} for verifiable training proposes rounding intermediate computations to eliminate the accumulation of errors, but results in high storage cost for the model provider who stores rounding decisions.

Other works (e.g.,~\cite{OngFerrante25}) take a heuristic approach, asserting that the numerical discrepancies between different hardware backends in typical ML workloads are ``small enough'' to be ignored. While such heuristics may hold in many practical scenarios, they lack formal guarantees, and it has not been shown whether they remain valid in adversarial or edge-case settings. For instance,  a malicious service provider might deliberately exploit non-determinism to introduce imperceptible but harmful changes. Therefore, developing a principled framework for \emph{verifiable and reproducible ML} that is robust to hardware-induced non-determinism remains a fundamental and urgent research challenge.

\paragraph{Our work.} 
Our main contribution is designing Hawkeye, a platform for \textbf{accurately reproducing matrix multiplication}, the fundamental operation behind training and inference, across hardware types without modifying the original
GPU kernels. Our only requirement is knowing which GPU was used for the execution.  We show that, despite architectural disparities, it is
possible to replicate the exact bit operations of NVIDIA's GPU Tensor Cores on CPUs.  Such a capability unlocks new verification workflows where an offline 
reference CPU run serves as an \emph{oracle} against which arbitrary
GPU executions can be checked at essentially no overhead. We test our framework across different GPU architectures and datatypes, showing a 100\% success rate in replicating large (\textit{4096 x 4096}) matrix multiplication. Source code for Hawkeye is provided at \url{https://github.com/badasherez/gpu-simulator}.

\section{The Challenge: Reproducing NNs}\label{sec:challenge}

At a very high level, the workflow of both training or running inference with a neural network (NN) consists primarily of a sequence of matrix multiplication (MatMul) operations interleaved with nonlinear element-wise functions, such as activation functions.

\paragraph{GPU architecture.}
Single-Instruction Multiple-Thread (SIMT) cores, often referred to as CUDA cores, serve as versatile computing units capable of executing a broad spectrum of instructions, such as integer arithmetic, floating-point calculations, and load/store operations. These cores process scalar or vector instructions over individual or grouped data elements. In contrast, Tensor Cores~\cite{nvidia_h100_2022,nvidia_a100_2020} are purpose-built hardware units optimized for high-throughput matrix multiplication. For example, the Tensor Cores on NVIDIA’s A100 and H100 GPUs deliver at least 10$\times$ improved performance for particular calculations over SIMT cores. Importantly, Tensor Cores operate on a coarser granularity:  a single \texttt{mma} instruction can multiply two $16 \times 16$ matrices  in one instruction.

\paragraph{Sources of non-determinism.}
Two main sources of non-determinism arise within this workflow due to the nature of floating-point arithmetic:

\begin{itemize}
    \item \textbf{Software-based non-determinism.} Floating-point operations performed in different orders across software implementations can yield varying numerical results, despite mathematical equivalence. This issue can be resolved by explicitly enforcing a consistent operation ordering, such as different algorithms for the convolution operation or batching ~\cite{HeumosECMMLGN23, pytorch_deterministic_algorithms, thinkingmachines_nondeterminism_llm}.
\item \textbf{Hardware-based non-determinism} Different hardware architectures can produce divergent floating-point results for the same computations due to hardware design choices. Libraries such as ~\cite{pyxis-roc_sass-math} accurately reproduce GPU element-wise functions on CPUs, but more complex operations such as matrix multiplication remain unresolved.
\end{itemize}

\paragraph{Problem statement.}

Given a specific GPU architecture employing Tensor Cores for matrix multiplication, we aim to replicate these Tensor Core computations precisely on CPU architectures, achieving bit-exact equivalence for every individual output. In particular, we seek to emulate the Tensor Core operation that updates an FP32 accumulator tile \(C\) with the product of two input tiles \(A\) and \(B\) of dimensions \(16 \times 16\) (in FP16, BF16, or FP8 precision), following
\[
C_{t+1} = C_t + A_t B_t .
\]

Achieving bit-exact equivalence is challenging because Tensor Core computations are implemented through hardware-specific mixed-precision pipelines whose numerical behavior is only partially exposed through the programming model. While individual floating-point operations follow IEEE~754 semantics, Tensor Cores use implementation-specific accumulation structures and internal intermediate representations. As a result, the effective accumulation order and rounding points differ from those produced by a straightforward CPU implementation, making faithful software emulation of Tensor Core arithmetic non-trivial.

\section{Floating-Point Preliminaries}

Floating-point (FP) numbers provide a method to represent real numbers using finite precision, characterized by three components: a \textsf{sign} bit ($s$), \textsf{exponent} bits ($e$), and \textsf{mantissa} (fractional) bits ($m$). According to the IEEE 754 standard~\cite{kahan1996ieee}, a FP number $x$ is represented as:
\begin{align*}
&        x = (-1)^{\textsf{sign}} \cdot 2^{\textsf{exponent} - \textsf{bias}} \cdot \textsf{mantissa} 
   \\
 &       \mathsf{bias}=2^{{len}(\textsf{exponent})-1} -1.
\end{align*}
 Additionally, the mantissa includes an implicit leading bit, typically 1 for normalized numbers and 0 for subnormal numbers.

Floating-point data types differ in their allocation of bits, and with Hawkeye we focus on the following types:

\begin{itemize}
\item \textbf{FP32 (Single precision)}: 1 sign bit, 8 exponent bits, and 23 mantissa bits. An FP32 number can represent values in the range of approximately $\pm 3.4 \times 10^{38}$.

\item \textbf{FP16 (Half precision)}: 1 sign bit, 5 exponent bits, and 10 mantissa bits. An FP16 number can represent numbers in the range $\pm 65,504$. 

\item \textbf{BF16 (Brain floating-point)}: 1 sign bit, 8 exponent bits, and 7 mantissa bits; exponent bias = 127.
A BF16 number shares the same exponent range as FP32 ($\pm 3.4 \times 10^{38}$), prioritizing dynamic range over precision.

\item \textbf{FP8 (E4M3 format)}:
1 sign bit, 4 exponent bits, and 3 mantissa bits; exponent bias = 7.
An FP8  number can represent values roughly in the range $\pm 448$. 

\end{itemize}

Finally, there exists exceptional bit patterns for floating point representations, including subnormal numbers (values closer to zero), infinity (all exponent bits are ones and the mantissa is zero), and NaN (all exponent bits are ones and the mantissa is non-zero). 
\subsection{Multiplication and Addition of FPs}
The IEEE 754 standard defines how floating-point  numbers are represented and manipulated in binary systems. It includes rules for operations like addition and multiplication to ensure consistency across different computing platforms.

We describe the algorithms for  floating-point multiplication and addition in  Algorithm~\ref{alg:fpmult} and Algorithm~\ref{alg:fpadd}, respectively. For clarity, the algorithms below omit special cases (e.g., NaN, infinity, subnormal numbers, overflow, and underflow), but present the core steps for normalized FP numbers.

\begin{breakablealgorithm}
\caption{IEEE 754 Floating-Point Multiplication}
\label{alg:fpmult}
\begin{algorithmic}[1]
\REQUIRE Floating-point inputs $(s_a, e_a, m_a)$ and $(s_b, e_b, m_b)$
\ENSURE Resulting floating-point value $(s_{\text{result}}, e_{\text{result}}, m_{\text{result}})$

\STATE \( s_\text{result} \gets s_a \oplus s_b \)
\STATE \( e_\text{result} \gets e_a + e_b - \text{bias} \)
\STATE \( m_\text{result} \gets (1 + m_a) \times (1 + m_b) \)
\IF{\( m_\text{result} \geq 2 \)}
    \STATE \( m_\text{result} \gets m_\text{result} >> 1 \)
    \STATE \( e_\text{result} \gets e_\text{result} + 1 \)
\ENDIF
\STATE Round \( m_\text{result} \) to the nearest representable number in the target FP data type, using tie-breaking towards even.
\end{algorithmic}
\end{breakablealgorithm}

\medskip
\begin{breakablealgorithm}
\caption{IEEE 754 Floating-Point Addition}
\label{alg:fpadd}
\begin{algorithmic}[1]
\REQUIRE Floating-point inputs $(s_a, e_a, m_a)$ and $(s_b, e_b, m_b)$
\ENSURE Resulting floating-point value $(s_{\text{result}}, e_{\text{result}}, m_{\text{result}})$

\STATE \( e_\text{max} \gets \max(e_a, e_b) \)
\STATE Align mantissas:
\[
m'_a \gets (1 + m_a) \times 2^{(e_a - e_\text{max})}, \quad m'_b \gets (1 + m_b) \times 2^{(e_b - e_\text{max})}
\]
\STATE Perform addition or subtraction based on signs:
\[
m_\text{result} \gets (-1)^{s_a} m'_a + (-1)^{s_b} m'_b
\]
\STATE \( s_\text{result} \gets \text{sign}(m_\text{result}) \)
\STATE \( m_\text{result} \gets |m_\text{result}| \)
\STATE \( e_\text{result} \gets e_\text{max} \)
\STATE Normalize \( m_\text{result} \):
\begin{itemize}
    \item If \( m_\text{result}\geq 2 \), shift \( m_\text{result} \) right until \( 1 \leq m_\text{result} < 2 \), incrementing \( e_\text{result} \) by the shift amount.
    \item If \( m_\text{result} < 1 \), shift \( m_\text{result} \) left until \( 1 \leq m_\text{result} < 2 \), decrementing \( e_\text{result} \) by the shift amount.
\end{itemize}
\STATE Remove implicit leading bit from \( m_\text{result} \) (retain fractional part only)
\STATE Round \( m_\text{result} \) to the nearest representable number in the target FP data type, using tie-breaking towards even.
\end{algorithmic}
\end{breakablealgorithm}

\paragraph{Non-associativity.}
Floating-point arithmetic is sensitive to the order of operations due to rounding. Although IEEE~754 specifies the exact result of each individual floating-point operation, different evaluation orders introduce different intermediate rounding steps, which can lead to different final results. For example, consider three FP16 values $a=65504$, $b=-65504$, and $c=1$. Evaluating $(a+b)+c$ yields $1$, since $a+b=0$ exactly and $0+1=1$. However, evaluating $a+(b+c)$ yields $0$: the sum $b+c=-65503$ cannot be represented in FP16 and rounds to $-65504$, after which $a+(-65504)=0$. 
\section{Overview of Hawkeye}

The goal of Hawkeye is to reproduce the bit-exact result of matrix multiplication with Tensor Cores on CPUs, in order to enable verifiable and reproducible machine learning. To this end, we developed a systematic testing methodology aimed at understanding and reproducing the internal numerical behavior of GPU Tensor Cores.

\paragraph{Tensor Core Characterization Methodology.}
To characterize the numerical behavior of Tensor Core matrix multiply-accumulate operations, we implement custom CUDA kernels that directly invoke the hardware MMA instructions via inline PTX assembly. For BF16 and FP16 on NVIDIA GPUs, the PTX \texttt{wmma.mma\_sync} instruction compiles to a single HMMA SASS instruction. Using these kernels, we execute a series of targeted tests designed to isolate individual properties of the internal accumulation pipeline.  These targeted tests are performed over the multiplication of three tiles, $A$, $B$, and $C$ (the accumulator), of dimensions 16 x 16 each, which we select to test certain properties.  Finally, we then encode these characterization results into a software simulator that reproduces the discovered accumulation semantics. 

\paragraph{Target Tests}
Hawkeye consists of a suite of targeted tests, each designed to characterize specific aspects of matrix multiply-accumulate (\texttt{mma}) operations on NVIDIA Tensor Cores. The same set of tests allows us to recover the exact sequence of operations across different GPU architectures and data types. While we expect the framework to generalize to additional architectures and precisions, our study focuses on three GPU architectures (Ampere, Hopper, and Ada Lovelace) and three precision formats (FP16, BF16, and FP8).

\begin{itemize}

\item \textbf{Summation Dependency and Order Test}: This test determines the precise sequence in which summations are executed during the accumulation phase.

\item \textbf{Internal Precision Detection Test}: This test identifies the exact internal numeric precision employed during each individual summation step within the Tensor Core computations.

\item \textbf{Rounding Mode Detection Test}: This test detects and characterizes the rounding mode applied for each arithmetic operation, enabling accurate replication of GPU rounding behavior.

\item \textbf{Normalization Stage Detection Test}: This test recovers at which specific stages the intermediate computational states are normalized into standard floating-point representations.

\item \textbf{Subnormal Behavior Detection Test}: This test characterizes the handling and representation of subnormal floating-point numbers throughout the Tensor Core computation pipeline.

\end{itemize}

By combining the results of these tests, we can reconstruct a computational model that accurately reproduces the Tensor Core arithmetic pipeline, ensuring bit-for-bit equivalence across every matrix multiplication output. In the following sections, we walk through the test outputs for reproducing multiplications of FP16 on Ampere (Section \ref{sec:fp16amp}), FP16 on Hopper (Section \ref{sec:fp16hopper}), and BFP16 on Ampere (Section \ref{sec:bfp16amp}). We found that Lovelace followed the same architecture structures as Ampere, and present empirical results confirming successful bit-wise replication over hundreds of thousands of randomly generated \(16 \times 16\) tiles in Section \ref{sec:results}.
\section{Reproducing FP16 Workflow on Ampere} \label{sec:fp16amp}
We now go through in detail how Hawkeye can enable replication of FP16 matrix multiplication on Ampere Tensor Cores. However, before applying our sequence of targeted tests, we will first verify that that a single multiplication of two FP16 numbers does retain full precision.

\subsection{Full-Precision FP16 Products Verification}
We first want to understand whether multiplication of two FP16 numbers  retains full precision in the context of a GPU-style dot product. To examine this behavior, We set \( A_{i,1} = B_{1,j} = 2^{11} - 1 \), which, results in a product that exceeds the dynamic range of FP16. This value produces a product that requires the widest mantissa—up to 21 bits—among all possible multiplications of two finite FP16 values, exceeding the precision directly representable in the FP16 format.

All other elements of tiles $A$ and $B$ are initialized to zero, and the accumulator $C_{i,j}$ is also set to zero. This configuration ensures that the dot-product reduces to a single multiplication: 
\[
D_{i,j} = C_{i,j} + A_{i,1} \cdot B_{1,j} = (2^{11} - 1)^2
\]

Note that the product $(2^{11} - 1)^2 = 4190209$  exceeds the range of FP16, and would therefore result in infinity if stored as FP16 during accumulation. However, after applying the Algorithm \ref{alg:fp16precision} Test, we see that the result is exact and no precision is lost. This is consistent with the known details of NVIDIA Tensor Cores \cite{nvidia_tensor_cores_cuda9}. 

\begin{breakablealgorithm}
\label{alg:fp16precision}
\caption{FP16$\times$FP16 Overflow and Precision Behavior}
\label{alg:fp16_product_precision}
\begin{algorithmic}[1]
\STATE Initialize tiles $A$, $B$, and $C$ to all zeros
\STATE Set $A_{0,1} \gets 2^{11} - 1 = 2047$
\STATE $B_{1,0} \gets 2^{11} - 1 = 2047$
\STATE Compute in the Tensor cores:
\[
D \gets C + A \cdot B 
\]
\STATE Verify $D_{0,0} = (2^{11} - 1)^2$ to ensure full precision product.
\end{algorithmic}
\end{breakablealgorithm}

\noindent\textbf{Conclusion}: A single product of two FP16 elements is maintained with full precision (FP32).

\subsection{Summation Dependency and Order}
We now proceed with the first test in Hawkeye: reverse-engineering the specific order of partial sums of products, which impacts the final bit-level result. This requires identifying computationally independent subgroups within the summation, which involves an  exhaustive search across all possible subsets of the 16 products. For each subset, we perform a call to Algorithm~\ref{alg:neutrality_test} to check for \textit{computational neutrality}. A subgroup is defined as computationally neutral if, when its elements are engineered to sum to zero, they have no impact on the final bit-level result of the total accumulation.

By collecting these neutral subgroups, we can reconstruct the nested structure of the calculation. This nesting shows how smaller groups of operations are bundled inside larger ones. For example, if a small set of products is found to be neutral and is also part of a larger neutral group, it indicates that the hardware first combines the smaller set into a partial sum before adding it to the rest of the elements. This hierarchy allows us to trace the branches of the hardware's summation logic, as well as determine whether the accumulator enters the pipeline at the beginning or the end of the process.
\\

\begin{breakablealgorithm}
\caption{Test for a Computationally Neutral Subgroup}
\label{alg:neutrality_test}
\begin{algorithmic}[1]
    \STATE \textit{S} is a set of product indices $\{k_1, k_2, ..., k_m\}$ from $\{0, ..., 16\}$.
    \STATE \textbf{Constants:}
    \STATE $V_{large} \gets 2^{20}$
    \STATE $V_{small} \gets 2^{-20}$
    \STATE
    \STATE \textbf{Test 1: Cancellation Scenario}
    \STATE $C_{0,0} \gets V_{large}$ if $0 \in S$ else $V_{small}$
    \FOR{$k \in \{1, ..., 16\}$}
        \IF{$k \notin S$}
            \STATE $A_{0,k},B_{k,0}\gets \sqrt{V_{small}},\sqrt{V_{small}}$
        \ELSE
            \STATE $A_{0,k},B_{k,0}\gets \sqrt{V_{large}},\sqrt{V_{large}}$
        \ENDIF
    \ENDFOR
 
    \STATE Multiply half of the elements in S by -1 in order to get zero sum
    \STATE $D\gets \text{Compute in Tensor cores}(C + AB)$
    \STATE $R_{cancel} \gets D_{0,0}$
    \STATE
    \STATE  \textbf{Test 2: Zeroed Subgroup Scenario (Baseline)}
    \STATE $C_{0,0} \gets 0$ if $0 \in S$ else $V_{small}$
    \FOR{$k \in \{1, ..., 16\}$}
        \IF{$k \notin S$}
            \STATE $A_{0,k},B_{k,0}\gets \sqrt{V_{small}},\sqrt{V_{small}}$
        \ELSE
            \STATE $A_{0,k},B_{k,0}\gets 0,0$
        \ENDIF
    \ENDFOR
    
    \STATE $D\gets \text{Compute in Tensor Cores}(C + AB)$
    \STATE $R_{zero} \gets D_{0,0}$
    \STATE \textbf{return} $\text{BitwiseCompare}(R_{cancel}, R_{zero})$
\end{algorithmic}
\end{breakablealgorithm}

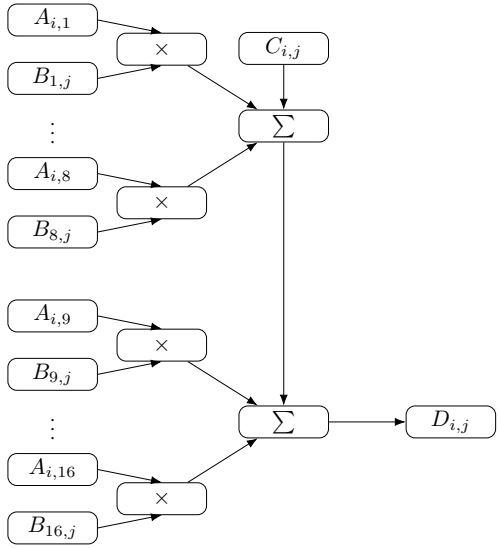
\begin{figure}[h!]
\centering
\begin{center}
\scalebox{0.85}{
\begin{tikzpicture}[
  >=Latex,
  node distance=4mm and 8mm,  
  box/.style={draw, rounded corners, align=center, inner sep=2pt, minimum width=14mm, minimum height=5mm},
  small/.style={font=\footnotesize}
]

\node[box] (A1) {$A_{i,1}$};
\node[box, below=of A1] (B1) {$B_{1,j}$};
\coordinate (M1) at ($(A1)!0.5!(B1)$);
\node[box, right=10mm of M1] (P1) {$\times$};
\node[small, below=1mm of B1] (dots18) {$\vdots$};

\node[box, below=1mm of dots18] (A8) {$A_{i,8}$};
\node[box, below=of A8] (B8) {$B_{8,j}$};
\coordinate (M8) at ($(A8)!0.5!(B8)$);
\node[box, right=10mm of M8] (P8) {$\times$};

\coordinate (MX1) at ($(P1)!0.5!(P8)$);
\node[box, right=12mm of MX1] (SUM1) {$\sum$};
\draw[-{Latex}] (A1.east) -- (P1.north);
\draw[-{Latex}] (B1.east) -- (P1.south);

\draw[-{Latex}] (A8.east) -- (P8.north);
\draw[-{Latex}] (B8.east) -- (P8.south);

\draw[-{Latex}] (P1) -- (SUM1);
\draw[-{Latex}] (P8) -- (SUM1);

\node[box, above=7mm of SUM1] (C) {$C_{i,j}$};
\draw[-{Latex}] (C) -- (SUM1);

\node[box, below=8mm of B8] (A9) {$A_{i,9}$};
\node[box, below=of A9] (B9) {$B_{9,j}$};
\coordinate (M9) at ($(A9)!0.5!(B9)$);
\node[box, right=10mm of M9] (P9) {$\times$};

\node[small, below=1mm of B9] (dots916) {$\vdots$};

\node[box, below=1mm of dots916] (A16) {$A_{i,16}$};
\node[box, below=of A16] (B16) {$B_{16,j}$};
\coordinate (M16) at ($(A16)!0.5!(B16)$);
\node[box, right=10mm of M16] (P16) {$\times$};
\coordinate (MX3) at ($(P9)!0.5!(P16)$);
\node[box, right=12mm of MX3] (SUM2) {$\sum$};

\draw[-{Latex}] (A9.east) -- (P9.north);
\draw[-{Latex}] (B9.east)  --  (P9.south);
\draw[-{Latex}] (A16.east) --  (P16.north);
\draw[-{Latex}] (B16.east) --  (P16.south);

\draw[-{Latex}] (P9) -- (SUM2);
\draw[-{Latex}] (P16) -- (SUM2);

\draw[-{Latex}] (SUM1) -- (SUM2);

\node[box, right=12mm of SUM2] (OUT) {$D_{i,j}$};
\draw[-{Latex}] (SUM2) -- (OUT);

\end{tikzpicture}
}
\end{center}
\caption{Computational graph of the two-stage tensor core accumulation in a pyramid structure. The initial accumulator $C_{i,j}$ and the first 8 products are summed into an intermediate result, which is then summed with the last 8 products.}
\label{fig:Ampere_comp_graph}
\end{figure}

\noindent \textbf{Conclusion}: 
Our experimental results show that the group containing the initial accumulator and the first eight products is the only non-singleton subgroup that is computationally neutral. We can conclude that the hardware implements the specific two-stage accumulation structure depicted in Figure~\ref{fig:Ampere_comp_graph}.
Crucially, there is no evidence of dynamic sorting or reordering within the summation mechanism, indicating a straightforward and deterministic accumulation strategy implemented in the  hardware.
\subsection{Inferring GPU Internal Representation During Summation}
We next design a test to infer the size of the internal significand (mantissa) used during summation. Specifically, our goal is to determine the smallest representable value that survives addition without being rounded, thereby revealing the effective precisionsof the internal accumulator.

In this test, formalized in Algorithm~\ref{alg:gpu_internal_repr}, we set $A_{i,1} = A_{i,2} = 1$, $B_{1,j} = 1$, and $B_{2,j} = -1$, so that not only are their combined contribution to the dot product equal to zero, but also so that any elements with significantly smaller exponents will be ``swallowed'' due to limited precision during addition. We then set a small, non-canceling, term by setting $A_{i,3} = 2^{\lceil -c/2 \rceil}$ and $B_{3,j} = 2^{\lfloor -c/2 \rfloor}$, which will result in the product $A_{i,3} B_{3,j} = 2^{-c}$. All other elements in $A$, $B$, $C$ were set to zero. We then gradually increase $c$ in order observe the smallest value of $c$ such that $D_{i,j} = 2^{-c}$ is preserved in the result, without being rounded out.

\begin{breakablealgorithm}
\caption{Reproducing GPU Internal Representation}
\label{alg:gpu_internal_repr}
\begin{algorithmic}[1]
\STATE Initialize matrices $A, B, C$ to zeros
\STATE Set $A_{i,1} \gets 1, \quad B_{1,j} \gets 1$
\STATE Set $A_{i,2} \gets 1, \quad B_{2,j} \gets -1$
\FOR{each integer $c \geq 1$}
\STATE Set $A_{i,3} \gets 2^{\lceil -c/2 \rceil}, \quad B_{3,j} \gets 2^{\lfloor -c/2 \rfloor}$
\STATE Compute in the Tensor cores:
\[
D \gets C + A \cdot B 
\]
\STATE \textbf{if} $D_{i,j} \ne 2^{-c}$ \textbf{then}
\STATE \quad Record $c-1$ as the smallest exponent surviving without rounding
\STATE \quad \textbf{break}
\STATE \textbf{end if}
\ENDFOR
\end{algorithmic}
\end{breakablealgorithm}

\noindent\textbf{Conclusion}: Running this experiment on NVIDIA Ampere Tensor Cores with FP16 datatype resulted in a smallest value of $c = 24$, indicating a 24-bit significand (including the implicit leading bit) consistent with FP32 accumulation.

\subsection{Rounding Mode in Shift Operation During Summation}

During floating-point addition, operands with different exponents must first be aligned by shifting the significand of the smaller-magnitude operand. This \emph{shift operation} may discard low-order bits when the shift exceeds the available precision of the internal accumulator. The handling of these discarded bits depends on the rounding mode used during alignment. In this section, we determine the rounding mode applied by Tensor Cores during this internal shift step.

\begin{breakablealgorithm} \caption{Recovering Rounding Mode} \label{alg:recover_rounding_mode} \begin{algorithmic}[1] \STATE Initialize matrices $A, B, C$ to zeros \STATE Set $A_{i,1}, A_{i,2} \gets 1$, $B_{1,j} \gets 1$, $B_{2,j} \gets -1$ \STATE \textbf{Test 1:} Set $A_{i,3} \gets 2^{-13}$, $B_{3,j} \gets 2^{-12}$ \STATE Compute dot product in the Tensor cores and verify result is 0. \STATE \textbf{Test 2:} Set $A_{i,3} \gets -2^{-13}$, keep $B_{3,j} \gets 2^{-12}$ \STATE Compute dot product in the Tensor and verify result is 0 (rules out \texttt{nearest\_tie\_towards\_minus\_infinity}, \texttt{towards\_minus\_infinity}) \STATE \textbf{Test 3:} Set $A_{i,3} \gets 2^{-12} + 2^{-13}$, $B_{3,j} \gets 2^{-12}$ \STATE Compute dot product in the Tensor and verify result is $2^{-24}$ (rules out \texttt{nearest\_tie\_towards\_even}, \texttt{towards\_even}) \STATE \textbf{Test 4:} Set $A_{i,3} \gets 2^{-13}$, $B_{3,j} \gets 2^{-12} + 2^{-13}$ \STATE Compute dot product in the Tensor and verify result is 0 (indicating truncation rather than rounding) \end{algorithmic} \end{breakablealgorithm}

To isolate this behavior, we construct dot products in which a very small value must be aligned against values of magnitude $1$ during accumulation. The formal procedure is shown in Algorithm~\ref{alg:recover_rounding_mode}. We initialize the matrices so that two products cancel exactly: $A_{i,1}=A_{i,2}=1$, $B_{1,j}=1$, and $B_{2,j}=-1$. This ensures that the mathematically correct result of the dot product is determined solely by a carefully chosen small term.

\paragraph{Test 1.}
We first set $A_{i,3}=2^{-13}$ and $B_{3,j}=2^{-12}$, producing the product $2^{-25}$. The observed result of the dot product is $0$. This outcome is consistent with several rounding strategies during the shift step, including nearest-based modes and directed rounding modes.

\paragraph{Test 2.}
Next, we set $A_{i,3}=-2^{-13}$ while keeping $B_{3,j}=2^{-12}$. The result remains $0$. If the shift operation used a directed rounding mode toward negative infinity, the discarded bits would produce a negative contribution, so we can rule out rounding modes biased toward $-\infty$.

\paragraph{Test 3.}
To test nearest-based rounding modes, we construct a product slightly larger than the halfway point by setting
$A_{i,3}=2^{-12}+2^{-13}$ and $B_{3,j}=2^{-12}$, producing
\[
(2^{-12}+2^{-13}) \cdot 2^{-12} = 2^{-24} + 2^{-25}.
\]
The observed result of the dot product is $2^{-24}$, which eliminates rounding strategies such as \texttt{nearest\_tie\_towards\_even} and \texttt{towards\_even}.

\paragraph{Test 4.}
Finally, we reverse the construction by setting $A_{i,3}=2^{-13}$ and $B_{3,j}=2^{-12}+2^{-13}$. This again produces a value slightly above the halfway threshold. The observed result is $0$, indicating that the extra bits introduced during alignment are discarded rather than rounded upward.

\paragraph{Conclusion.}
Across these experiments, the observed behavior is consistent with a rounding mode of \texttt{towards\_zero} during the internal shift operation. In other words, when the alignment step discards low-order bits, Tensor Cores truncate these bits rather than rounding them to the nearest representable value.

\subsection{Post-Multiplication Normalization}

We next investigate whether intermediate products are normalized before being forwarded to the accumulation stage. Normalization means adjusting a number’s representation so that the significand (mantissa) falls within a standard range. An alternative design is to defer normalization and pass the raw product significand directly to the accumulator.

To determine which behavior Tensor Cores implement, we construct a dot product that produces large intermediate products that cancel exactly while introducing a very small third term (Algorithm~\ref{alg:recover_normalization_overflow}). Specifically, we set
$A_{i,1}=A_{i,2}=1.5$, $B_{1,j}=1.5$, and $B_{2,j}=-1.5$, yielding products of $2.25$ and $-2.25$ that cancel during accumulation. We then introduce a small term by setting $A_{i,3}=2^{-12}$ and $B_{3,j}=2^{-12}$, producing the product $2^{-24}$.

If products were normalized immediately after multiplication, the renormalization step would increase the exponent of the large intermediate values (e.g., $2.25 = 1.125 \times 2^1$), increasing the exponent difference during accumulation. This larger alignment shift could cause the small value $2^{-24}$ to be discarded during exponent alignment. Conversely, if normalization is deferred, the raw product significand is accumulated directly, preserving the small contribution.

\begin{breakablealgorithm}
\caption{Detecting Post-Multiplication Normalization}
\label{alg:recover_normalization_overflow}
\begin{algorithmic}[1]
\STATE Initialize matrices $A, B, C$ to zeros
\STATE Set $A_{i,1}, A_{i,2} \gets 1.5$, $B_{1,j} \gets 1.5$, $B_{2,j} \gets -1.5$
\STATE Set $A_{i,3} \gets 2^{-12}, \quad B_{3,j} \gets 2^{-12}$
\STATE Compute in the Tensor cores:
\[
D \gets C + A \cdot B
\]
\STATE Check whether $D_{i,j} = 2^{-24}$
\end{algorithmic}
\end{breakablealgorithm}

With Amper FP16, we observed $D_{i,j}=2^{-24}$, indicating that the small term survives accumulation. This behavior is consistent with a design in which intermediate products are not normalized prior to summation. Instead, the raw multiplication result is forwarded directly to the accumulation pipeline, allowing the internal product significand to temporarily exceed the normalized range.

\subsection{Normalization After Subnormal Multiplication}

We next investigate how Tensor Cores handle products involving subnormal inputs. In IEEE floating-point arithmetic, subnormal numbers do not contain the implicit leading 1 in the significand, which effectively reduces the available precision. Some floating-point pipelines renormalize such values during multiplication, restoring a normalized representation before forwarding the result to the accumulation stage. Alternatively, the hardware may propagate the reduced-precision significand directly into the accumulator.

To distinguish between these behaviors, we construct a test in which a subnormal value is multiplied by a large normal value and then combined with a smaller term whose preservation depends on the effective precision of the intermediate product (Algorithm~\ref{alg:recover_subnormal_normalization}).

Specifically, we set $A_{i,1} = 2^{-14-k}$, which is a subnormal FP16 value, and $B_{1,j} = 2^{14}$. Their product equals
\[
2^{-14-k} \cdot 2^{14} = 2^{-k}.
\]
Although the resulting value is normal, its significand originates from a subnormal input and therefore contains only $k$ significant bits of precision. We then introduce a smaller value by setting $A_{i,2} = 2^{-24}$ and $B_{2,j} = 1$, allowing us to observe whether this contribution survives the subsequent accumulation.

\begin{breakablealgorithm}
\caption{Detecting Normalization After Subnormal Multiplication}
\label{alg:recover_subnormal_normalization}
\begin{algorithmic}[1]
\STATE Initialize matrices $A, B, C$ to zeros
\STATE Set $A_{i,1} \gets 2^{-14-k}$ (subnormal FP16 value)
\STATE Set $B_{1,j} \gets 2^{14}$ (normal FP16 value)
\STATE Set $A_{i,2} \gets 2^{-24}, \quad B_{2,j} \gets 1$
\STATE Compute in the Tensor cores:
\[
D \gets C + A \cdot B
\]
\STATE Check whether the term $2^{-24}$ is preserved in $D_{i,j}$
\end{algorithmic}
\end{breakablealgorithm}

\noindent
\textbf{Conclusion.}
We observe that the product derived from the subnormal operand behaves as if its significand contains only the reduced precision inherited from the input. The smaller term is therefore more easily lost during alignment in the accumulation stage. This indicates that Tensor Cores do not renormalize the result of a subnormal multiplication prior to accumulation; instead, the reduced-precision significand produced by the multiplication is propagated directly into the summation pipeline.
\subsection{Rounding Mode of the Final Summation Result}

Finally, we determine the rounding mode applied when the internal accumulation result is converted to the output precision. Earlier experiments showed that the Tensor Core accumulator maintains higher precision than the target FP16 format. When the final result is written back, excess mantissa bits must therefore be discarded or rounded.

To determine how this conversion is performed, we construct a dot product in which a very large intermediate value is perturbed by a small correction term (Algorithm~\ref{alg:rounding_mode_final_result}). The large product establishes the dominant magnitude of the result, while the smaller term introduces low-order bits whose treatment reveals the rounding policy.

\paragraph{Test 1 (baseline magnitude).}
We first set
\[
A_{i,1} = B_{1,j} = 1.5 \cdot 2^{12}, \quad
A_{i,2} = 3, \quad B_{2,j} = 1 .
\]
The dominant product is
\[
(1.5 \cdot 2^{12})^2 = 2.25 \cdot 2^{24},
\]
while the smaller product contributes $3$. The observed result is
\[
D_{i,j} = 2.25 \cdot 2^{24},
\]
indicating that the small contribution is eliminated during the final precision reduction.

\paragraph{Test 2 (sign-changing correction).}
Next, we replace the correction term with
\[
A_{i,2} = 1, \quad B_{2,j} = -1 ,
\]
producing a product of $-1$. The observed result becomes
\[
D_{i,j} = 2.25 \cdot 2^{24} - 5.
\]
The magnitude of the adjustment corresponds to the truncation of low-order mantissa bits rather than rounding to the nearest representable value.

\begin{breakablealgorithm}
\caption{Recovering Rounding Mode of Final Result}
\label{alg:rounding_mode_final_result}
\begin{algorithmic}[1]
\STATE Initialize matrices $A,B,C$ to zero

\STATE \textbf{Test 1:}
\STATE $A_{i,1} \gets 1.5 \cdot 2^{12}$,\quad $B_{1,j} \gets 1.5 \cdot 2^{12}$
\STATE $A_{i,2} \gets 3$,\quad $B_{2,j} \gets 1$
\STATE Compute $D \gets C + A \cdot B$

\STATE \textbf{Reset matrices}

\STATE \textbf{Test 2:}
\STATE $A_{i,1} \gets 1.5 \cdot 2^{12}$,\quad $B_{1,j} \gets 1.5 \cdot 2^{12}$
\STATE $A_{i,2} \gets 1$,\quad $B_{2,j} \gets -1$
\STATE Compute $D \gets C + A \cdot B$
\end{algorithmic}
\end{breakablealgorithm}

\noindent
\textbf{Conclusion.}
The observed behavior indicates that the final conversion from the internal accumulator representation to the output format uses \texttt{towards\_zero} rounding. In other words, excess mantissa bits are truncated rather than rounded to the nearest representable value.

\subsection{Recovered Tensor Core Computation Pipeline}

Based on the experiments described in the previous sections, we reconstruct the numerical pipeline used by NVIDIA Ampere Tensor Cores for FP16 matrix multiply-accumulate operations. The resulting model captures the sequence of transformations applied to each partial product and the accumulation strategy used to produce the final output.

Our findings indicate that Tensor Cores follow a structured pipeline consisting of three stages: (1) multiplication with expanded internal precision, (2) grouped accumulation with truncation-based exponent alignment, and (3) final normalization and precision reduction to the output format. The recovered computational model is summarized in Algorithms~\ref{alg:ampere_fp16_mult},~\ref{alg:ampere_fp16_summation}, and~\ref{alg:ampere_fp16_dotproduct}, with the corresponding computational graph illustrated in Figure~\ref{fig:Ampere_comp_graph}.
This recovered model captures the key numerical behaviors observed in our experiments, including delayed normalization, truncation-based alignment during summation, and the hierarchical grouping structure used for accumulation.
\section{Reproducing FP16 Workflow on Hopper}  \label{sec:fp16hopper}
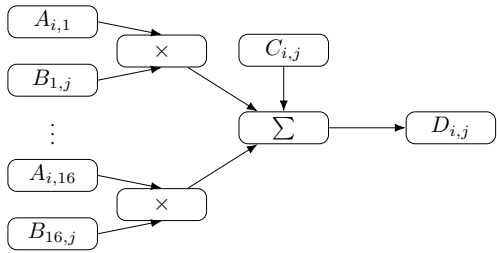
\begin{figure}[h!]
\centering
\begin{center}
\scalebox{0.85}{
\begin{tikzpicture}[
  >=Latex,
  node distance=4mm and 8mm,  
  box/.style={draw, rounded corners, align=center, inner sep=2pt, minimum width=14mm, minimum height=5mm},
  small/.style={font=\footnotesize}
]

\node[box] (A1) {$A_{i,1}$};
\node[box, below=of A1] (B1) {$B_{1,j}$};
\coordinate (M1) at ($(A1)!0.5!(B1)$);
\node[box, right=10mm of M1] (P1) {$\times$};
\node[small, below=1mm of B1] (dots18) {$\vdots$};

\node[box, below=1mm of dots18] (A8) {$A_{i,16}$};
\node[box, below=of A8] (B8) {$B_{16,j}$};
\coordinate (M8) at ($(A8)!0.5!(B8)$);
\node[box, right=10mm of M8] (P8) {$\times$};

\coordinate (MX1) at ($(P1)!0.5!(P8)$);
\node[box, right=12mm of MX1] (SUM1) {$\sum$};
\draw[-{Latex}] (A1.east) -- (P1.north);
\draw[-{Latex}] (B1.east) -- (P1.south);

\draw[-{Latex}] (A8.east) -- (P8.north);
\draw[-{Latex}] (B8.east) -- (P8.south);

\draw[-{Latex}] (P1) -- (SUM1);
\draw[-{Latex}] (P8) -- (SUM1);

\node[box, above=7mm of SUM1] (C) {$C_{i,j}$};
\draw[-{Latex}] (C) -- (SUM1);


\node[box, right=12mm of SUM1] (OUT) {$D_{i,j}$};
\draw[-{Latex}] (SUM1) -- (OUT);

\end{tikzpicture}
}
\end{center}
\caption{Computational graph of Hopper Tensor Core accumulation in a pyramid structure. The initial accumulator $C_{i,j}$ and the 16 products are summed into the final result}
\label{fig:hopper_comp_graph}
\end{figure}

To extend our analysis, we applied the same computational neutrality tests to the Hopper architecture Tensor Cores. The results revealed a different, more unified accumulation strategy. In contrast to the two-stage process found in the Ampere architecture, our tests on Hopper show that all 16 products are summed together with the initial accumulator in a single stage, as depicted in Figure~\ref{fig:hopper_comp_graph}.

Furthermore, our analysis indicates that the Hopper architecture utilizes an internal mantissa representation of 25 bits for this accumulation compared to 24 in Ampere. Consistent with the Ampere architecture, the final sum is normalized only after accumulation is complete, and the rounding mode is round-towards-zero. 
\section{Reproducing BF16 Workflow on Ampere} \label{sec:bfp16amp}

We now extend our investigation of the dot product behavior to the BF16 format. Initial experiments show that all previously observed behaviors in FP16, including summation order, accumulator integration, rounding modes, and normalization patterns, are preserved when operating in BF16 on Ampere GPUs. This confirms that the underlying dot product algorithm is format-agnostic in terms of execution strategy.

However, BF16 introduces two new cases that do not occur in FP16-based tile multiplications. First, due to its wide exponent range, multiplications within a tile can result in products that fall below the smallest normalized FP32 value, producing sub-normal results in the accumulation path. Second, it is also possible for intermediate products or accumulations to exceed the dynamic range of FP32, resulting in overflow beyond the representable range of the FP32 format. These scenarios require additional tests to understand how such edge cases are handled, including whether rounding, saturation, or flushing to zero is applied by the hardware.

\subsection{Handling of Out-of-Range Values in Tile Multiplication}

To investigate how the GPU handles intermediate values that exceed the representable FP32 range during BF16 tile multiplication, we conducted two targeted tests. In the first test, we set $A_{i,1} = A_{i,2} = A_{i,3} = 2^{127}$ and $B_{1,j} = 2$, $B_{2,j} = -2$, $B_{3,j} = 1$. The expected contribution from the first two terms cancels out, and the result is $D_{i,j} = 2^{127}$. the intermediate values exceeds the maximum finite FP32 number ($\approx 3.4 \times 10^{38}$), yet the output was correctly returned as $2^{127}$, indicating that the hardware internally allows intermediate values to temporarily exceed the FP32 range during summation, as long as the final result falls back within the FP32 representable range.

In the second test, we set $A_{i,1} = A_{i,9} = 2^{127}$, with $B_{1,j} = 2$ and $B_{9,j} = -2^{127}$. The resulting dot product was $D_{i,j} = \infty$, suggesting that the intermediate result exceeded the FP32 dynamic range and was ultimately cast to infinity. This confirms that after the accumulation step, the final result is cast back to FP32, and any value that falls outside the FP32 representable bounds is mapped to positive or negative infinity according to the IEEE 754 standard.

\begin{algorithm}
\caption{Handling Out-of-Range in BF16}
\label{alg:handling_bf16_oor}
\begin{algorithmic}[1]
\STATE \textbf{Require:} Matrices $A,B \in \mathbb{R}^{16 \times 16}$, accumulator $C \in \mathbb{R}^{16 \times 16}$
\STATE Initialize all entries of $A,B,C$ to $0$

\STATE \textbf{Test 1 (temporary extended-range accumulation):}
\STATE $A_{i,1}, A_{i,2}, A_{i,3} \gets 2^{127}$ \hfill\textit{(BF16 inputs)}
\STATE $B_{1,j} \gets 2,\quad B_{2,j} \gets -2,\quad B_{3,j} \gets 1$
\STATE Compute in the Tensor cores:
\[
D \gets C + A \cdot B 
\]
\STATE \textbf{Verify:} $D_{i,j} = 2^{127}$ \hfill\textit{(final FP32 within range despite intermediate overflow)}

\STATE \textbf{Reset:} Set all entries of $A,B,C$ back to $0$

\STATE \textbf{Test 2 (final cast to FP32 with overflow):}
\STATE $A_{i,1}, A_{i,9} \gets 2^{127}$ \hfill\textit{(BF16 inputs)}
\STATE $B_{1,j} \gets 2,\quad B_{9,j} \gets -2^{127}$
\STATE Compute in the Tensor cores:
\[
D \gets C + A \cdot B 
\]
\STATE \textbf{Verify:} $D_{i,j} = \infty$ \hfill\textit{(magnitude exceeds FP32; saturated to IEEE 754 infinity)}
\end{algorithmic}
\end{algorithm}

\noindent\textbf{Conclusion}:
These results indicate that while the GPU supports extended-range accumulation internally when using BF16, the final output still conforms to FP32 limits, and extreme overflows are handled via saturation to infinity.

\subsection{Minimal Effective Contribution from Sub-Normal Products}

As established in previous sections, the GPU performs internal summation using an extended exponent range beyond the standard FP32 format. This higher internal precision allows the accumulation of intermediate values with exponents far outside the final representable FP32 range. Motivated by this behavior, we tested the smallest multiplicative contribution that can still affect the final dot-product result. Since the internal summation uses rounding towards zero, the only way a small product can influence the result is by shifting into the least significant bits of the FP32 accumulation and reducing a bit that is still within the representable range.

We found that the smallest effective contribution corresponds to a product with exponent $2^{-156}$. This implies that the summation logic tracks enough significant bits to detect changes at this level. For group-based alignment, this threshold is equivalent to taking the maximum exponent across all products and the number $-132$ (For Hopper GPUs, the corresponding number is $ -133$). To generalize this observation, we designed a reproducible test framework that identifies the minimal maximal exponent during accumulation across architectures, formalized in Algorithm~\ref{alg:gpu_minmax_exp}.

To confirm this, we ran the following test: we set $A_{i,1} = A_{i,2} = B_{j,1} = 2^{-74}$, and $B_{2,j} = -2^{-82}$. The result of the dot product was $D_{i,j} = 2^{-149}$. However, when we modified $B_{2,j}$ to $-2^{-83}$, the result increased to $D_{i,j} = 2^{-148}$. This change confirms that a contribution at the $2^{-156}$ level still affects rounding behavior due to its interaction with the mantissa bits of the aligned sum. Any smaller contribution would be fully truncated and have no effect.
\begin{breakablealgorithm}
\caption{Reproducing GPU Minimal max exponent}
\label{alg:gpu_minmax_exp}
\begin{algorithmic}[1]
\STATE \textbf{Input:} $m$ - GPU  internal representation
\STATE Initialize matrices $A, B, C$ to zeros
\FOR{each integer $e \leq 0$}
\STATE Set $A_{i,1} \gets 2^{\lceil \frac{e}{2} \rceil}, \quad B_{1,j} \gets 2^{\lfloor \frac{e}{2} \rfloor}$
\STATE Set $A_{i,2} \gets 2^{\lceil \frac{e-m}{2} \rceil}, \quad B_{2,j} \gets 2^{\lfloor \frac{e-m}{2} \rfloor}$
\STATE Compute in the Tensor cores:
\[
D \gets C + A \cdot B 
\]
\STATE \textbf{if} $D_{i,j} = 2^{e}$ \textbf{then}
\STATE \quad Record $e+1$ as the minimal option for the maximal exponent during accumulation.
\STATE \quad \textbf{break}
\STATE \textbf{end if}
\ENDFOR
\end{algorithmic}
\end{breakablealgorithm}

\section{Empirical Results}\label{sec:results}

Recall that the primary motivation of Hawkeye is verifiable machine learning: given a server's computation on an arbitrary GPU, can an auditor replicate the computation such that any discrepancies must be due to incorrect behavior versus GPU nondeterminism? Due to scope of our paper, we leave integrating matrix multiplication implementations identified by Hawkeye into the architecture of an existing model for future work.  However, we provide simulators across the different GPU architectures and precision types we study in a public repository available here: \url{https://github.com/badasherez/gpu-simulator}.

To test our simulators, we create bit-exact reproduction tests across architectures (e.g. H100) and precision types (e.g. FP16), even extending our prior analysis to FP8 (E4M3 type) precision. These tests compare the outputs of our simulator with outputs of the custom kernels invoking \texttt{mma} on the hardware the simulator is run on. Across 100,000 randomly generated 16x16 matrix multiplication, our tests demonstrate 100\% bit-exact replication. 

We report the average CPU execution times for sample matrix multiplications in Table \ref{tab:times}. Our implementation provides a performance baseline, and further optimizations (such as a integer-based GPU  implementation of the simulators) are deferred to future work. However, for verifiable ML, it suffices for an auditor to reproduce a computation only once, and therefore a slow CPU-side execution can be tolerated.

\section{Future Work and Conclusion}

While Hawkeye demonstrates that GPU-level numerical behaviors can be faithfully reproduced on CPUs (or other GPU hardwares), several challenges remain before such techniques can be fully integrated into verifiable machine learning pipelines, including proof systems and distributed training settings. First, our current framework focuses on a subset of NVIDIA architectures that utilize Tensor Cores; extending this approach to other hardware vendors and accelerator designs is an important direction for future work. Although vendor-specific implementations may introduce subtle differences, we believe that our test suite captures the fundamental design choices underlying $16 \times 16$ tiled matrix multiplication, providing a strong basis for generalization.

Second, while Hawkeye accurately reproduces low-level matrix multiplication behavior, higher-level operations commonly used in modern ML workloads—such as convolutions or attention mechanisms —require additional reverse engineering of implementation details beyond the scope of this work. Bridging this gap is essential for enabling end-to-end reproducibility of full training and inference pipelines. 

Finally, integrating Hawkeye into practical verification frameworks raises systems-level challenges, including scalability in distributed environments and compatibility with cryptographic proof systems. Addressing these challenges will be critical for deploying Hawkeye in real-world auditing and verification workflows.
Overall, Hawkeye provides a foundation for understanding and reproducing hardware-level numerical behavior, paving the way toward reproducible and hardware-agnostic machine learning systems.
\section*{Acknowledgements}
Special thanks to Itamr Schen and Yoray Herzberg for their technical assistance in writing and implementing the CUDA kernels used in this work.

\bibliographystyle{plainnat}   
\bibliography{refs}  
\newpage
\appendix

\begin{breakablealgorithm}
\caption{Floating-Point Multiplication (Recovered Ampere Model for FP16)}
\label{alg:ampere_fp16_mult}
\begin{algorithmic}[1]
\REQUIRE Floating-point inputs $(s_A,e_A,m_A)$ and $(s_B,e_B,m_B)$
\ENSURE Product $(s_{\text{result}},e_{\text{result}},m_{\text{result}})$

\STATE $s_{\text{result}} \gets s_A \oplus s_B$
\STATE $e_{\text{result}} \gets e_A + e_B$
\STATE $m_{\text{raw}} \gets m_A \cdot m_B$
\STATE $m_{\text{result}} \gets m_{\text{raw}} \ll 3$ \hfill // internal precision expansion
\RETURN $(s_{\text{result}},e_{\text{result}},m_{\text{result}})$
\end{algorithmic}
\end{breakablealgorithm}

\begin{breakablealgorithm}
\caption{Grouped Summation (Recovered Ampere Model for FP16)}
\label{alg:ampere_fp16_summation}
\begin{algorithmic}[1]
\REQUIRE Floating-point values $\{(s_i,e_i,m_i)\}_{i=1}^{n}$
\ENSURE Result $(s_{\text{out}},e_{\text{out}},m_{\text{out}})$

\FOR{$i=1$ to $n$}
    \STATE $m_i \gets m_i \ll 1$ \hfill // expand to internal precision
\ENDFOR

\STATE $e_{\max} \gets \max_i e_i$

\FOR{$i=1$ to $n$}
    \STATE $\Delta e \gets e_{\max}-e_i$
    \STATE $m_i \gets m_i \gg \Delta e$ \hfill // truncating alignment shift
\ENDFOR

\STATE $M \gets \sum_{i=1}^{n} (-1)^{s_i} \cdot m_i$

\STATE Normalize $(e_{\max},M)$ to floating-point form
\STATE Truncate significand to output precision

\RETURN $(s_{\text{out}},e_{\text{out}},m_{\text{out}})$
\end{algorithmic}
\end{breakablealgorithm}

\begin{breakablealgorithm}
\caption{Dot Product Computation (Recovered Ampere Model for FP16)}
\label{alg:ampere_fp16_dotproduct}
\begin{algorithmic}[1]
\REQUIRE Accumulator $C$ and vectors $A[1..16], B[1..16]$
\ENSURE Dot product result

\FOR{$i=1$ to $16$}
    \STATE $P[i] \gets \textbf{Multiply}(A[i],B[i])$
\ENDFOR

\STATE $G_1 \gets \text{GroupSum}(\{C,P[1],\dots,P[8]\})$
\STATE $G_2 \gets \text{GroupSum}(\{G_1,P[9],\dots,P[16]\})$

\RETURN  $ G_2$
\end{algorithmic}
\end{breakablealgorithm}

\begin{table}[t]
\caption{Average execution time (in seconds) and standard deviatation for $4096 \times 4096$ matrix multiplication on an Apple M4 Pro CPU, calculated over 10 independent runs.}
\label{tab:times}
\vskip 0.15in
\begin{center}
\begin{small}
\begin{sc}
\begin{tabular}{lcccr}
\toprule
Precision & Architecture  & Time (sec.) & std (sec.) \\
\midrule
FP16    &  Ampere & 50.8 & 3.2 \\
FP16    &  Hopper  & 47.2 & 2.5 \\
BF16    &  Ampere  & 52.5 & 2.9 \\
BF16    &  Hopper  & 48.2 & 2.6 \\
FP8   &  Hopper  & 40.6 & 0.6 \\
\bottomrule
\end{tabular}
\end{sc}
\end{small}
\end{center}
\vskip -0.1in
\end{table}

\end{document}